\documentclass[preprint,eqsecnum,aps,nofootinbib]{revtex4}
\usepackage{amsfonts,amsmath,amssymb,amsthm}
\usepackage{latexsym}
\usepackage{bbm,bm}
\usepackage{graphicx}

\newcommand{\beq}{\begin{equation}}
\newcommand{\eeq}{\end{equation}}
\newcommand{\beqs}{\begin{eqnarray}}
\newcommand{\eeqs}{\end{eqnarray}}

\begin{document}

\title{ Quantum Entanglement with Generalized Uncertainty Principle}

\author{DaeKil Park$^{1,2}$\footnote{dkpark@kyungnam.ac.kr} }

\affiliation{$^1$Department of Electronic Engineering, Kyungnam University, Changwon
                 631-701, Korea    \\
             $^2$Department of Physics, Kyungnam University, Changwon
                  631-701, Korea    
                      }

\begin{abstract}
We explore how the quantum entanglement is modified in the generalized uncertainty principle (GUP)-corrected quantum mechanics by introducing the coupled harmonic oscillator system.
Constructing the ground state $\rho_0$ and its reduced substate $\rho_A = \mbox{Tr}_B \rho_0$, we compute two entanglement measures of $\rho_0$, i.e.  ${\cal E}_{EoF} (\rho_0) = S_{von} (\rho_A)$ 
and ${\cal E}_{\gamma} (\rho_0) = S_{\gamma} (\rho_A)$, where $S_{von}$ and $S_{\gamma}$ are the von Neumann and R\'{e}nyi entropies, up to the first order of the GUP parameter $\alpha$. It is shown that 
${\cal E}_{\gamma} (\rho_0)$ increases with increasing $\alpha$ when $\gamma = 2, 3, \cdots$. The remarkable fact is that ${\cal E}_{EoF} (\rho_0)$ does not have first-order of $\alpha$.
Based on there results we conjecture that ${\cal E}_{\gamma} (\rho_0)$ increases or decreases with increasing $\alpha$ when $\gamma > 1$ or $\gamma < 1$ respectively for nonnegative real $\gamma$. 
\end{abstract}

\maketitle

\section{Introduction}

As IC (integrated circuit) becomes smaller and smaller in modern classical technology, the effect of quantum mechanics becomes prominent more and more. As a result, quantum technology (technology 
based on quantum mechanics and quantum information theories\cite{text}) becomes important more and more recently. The representative constructed by quantum technology
is a quantum computer\cite{supremacy-1}, which was realized recently by making use of superconducting qubits. 
In quantum information processing quantum entanglement\cite{text,schrodinger-35,horodecki09} plays an important role as a physical resource. 
It is used in various quantum information processing, such as  quantum teleportation\cite{teleportation,Luo2019},
superdense coding\cite{superdense}, quantum cloning\cite{clon}, quantum cryptography\cite{cryptography,cryptography2}, quantum
metrology\cite{metro17}, and quantum computer\cite{supremacy-1,qcreview,computer}. Furthermore, with many researchers trying to realize such quantum information processing in the laboratory for the last few decades, quantum cryptography and quantum computer seem to approaching the commercial level\cite{white,ibm}.

Physics at the Planck scale suggests the existence of the minimal length (ML).  
The existence of the ML at this scale seems to be a universal characteristic of quantum gravity\cite{townsend76,amati89,garay94}. It appears in loop quantum gravity\cite{rovelli88,rovelli90,rovelli98,carlip01},
string theory\cite{konishi90,kato90,strominger91}, path-integral quantum gravity\cite{padmanabhan85,padmanabhan85-2,padmanabhan86,padmanabhan87,greensite91}, and black hole physics\cite{maggiore93}. 
ML also appeared in some microscope thought-experiment\cite{mead64}. 
From an aspect of quantum mechanics the existence of ML modifies the uncertainty principle from Heisenberg uncertainty principle (HUP)\cite{uncertainty,robertson1929}
$\Delta P \Delta Q \geq \frac{\hbar}{2}$ 
to generalized uncertainty principle (GUP)\cite{kempf93,kempf94}. This is because of the fact that the uncertainty of the position $\Delta Q$ should be larger than the ML. 

Then, it is natural to ask how the quantum entanglement is modified at the Planck scale. 
This question might be important to unveil the role of the quantum information at the Planck scale or early universe.  
In order to explore this issue we choose the GUP as
the simplest form proposed in Ref. \cite{kempf94}:
\begin{equation}
\label{GUP-d-1}
\Delta P_i \Delta Q_i \geq \frac{\hbar}{2} \left[ 1 + \alpha \left\{ (\Delta {\bf P})^2 + \langle \widehat{{\bf P}} \rangle^2 \right\} + 2 \alpha \left\{ (\Delta P_i)^2 + \langle \widehat{P}_i \rangle^2 \right\} \right]
\hspace{.7cm}  (i = 1, 2, \cdots, d)
\end{equation}
where $\alpha$ is a GUP parameter, which has a dimension $(\mbox{momentum})^{-2}$. Using 
$\Delta A \Delta B \geq \frac{1}{2} | \langle [\widehat{A}, \widehat{B}] \rangle |$, Eq. (\ref{GUP-d-1}) induces the 
modification of the commutation relation as 
\begin{eqnarray}
\label{GUP-d-2}
&& \left[ \widehat{Q}_i, \widehat{P}_j \right] = i \hbar \left( \delta_{ij} + \alpha \delta_{ij} \widehat{{\bf P}}^2 + 2 \alpha \widehat{P}_i \widehat{P}_j  \right)    \\    \nonumber
&& \hspace{1.0cm} \left[ \widehat{Q}_i, \widehat{Q}_j \right] = \left[\widehat{P}_i, \widehat{P}_j \right] = 0.
\end{eqnarray}

The existence of the ML can be seen in Eq. (\ref{GUP-d-1}). If $\langle \widehat{{\bm P}} \rangle = 0$ for simplicity, the equality of Eq. (\ref{GUP-d-1}) yields
\begin{equation}
\label{minimal-length}
\Delta Q_i^2 \geq \Delta Q_{i,min}^2 = 3 \alpha \hbar^2  \hspace{.5cm}  (i = 1, 2, \cdots, d)
\end{equation}
which arises when $\Delta P_j = 0 \hspace{.2cm}  (j \neq i)$.
If $\alpha$ is small, Eq. (\ref{GUP-d-2}) can be solved as 
\begin{equation}
\label{GUP-d-4}
\widehat{P}_i = \widehat{p}_i \left(1 + \alpha \widehat{{\bf p}}^2 + \alpha^2 \widehat{{\bf p}}^4 \right) + {\cal O} (\alpha^3)   \hspace{1.0cm} 
\widehat{Q}_i = \widehat{q}_i \left(1 + \alpha^2 \widehat{{\bf p}}^4 \right) + {\cal O} (\alpha^3)
\end{equation}
where $p_i$ and $q_i$ obey the usual Heisenberg algebra $[q_i, p_j] = i \hbar \delta_{ij}$. Thus, the ordering ambiguity occurs at ${\cal O} (\alpha^2)$.
We will use Eq. (\ref{GUP-d-4}) in the following to compute the quantum entanglement within the 
first order of $\alpha$. 

As commented before the purpose of this paper is to examine how the quantum entanglement is modified in the GUP-corrected quantum mechanics.  
In order to explore the issue we consider the two harmonic oscillator systems, which are coupled with each other via the 
quadratic term. The Hamiltonian of the system is presented in section II. In section III we derive the vacuum state $\rho_0$ and its reduced substate $\rho_A$. In this paper we adopt the entanglement measure for 
$\rho_0$ as von Neumann and R\'{e}nyi entropies of the substate;
\begin{eqnarray} 
\label{emeasure1}
&& {\cal E}_{EoF} (\rho_0) = S_{von} (\rho_A) = - \mbox{Tr} (\rho_A \ln \rho_A)                    \\     \nonumber
&& {\cal E}_{\gamma} (\rho_0) = S_{\gamma} (\rho_A) = \frac{1}{1 - \gamma} \ln \left( \mbox{Tr} \rho_A^{\gamma} \right),
\end{eqnarray}
where $S_{von}$ and $S_{\gamma}$ denote the von Neumann and R\'{e}nyi entropies. It is easy to show that these entanglement measures are invariant in the choice of the substate due to the Schmidt decomposition\cite{text}.
The first entanglement measure is the most popular one called ``entanglement of formation (EoF)''\cite{benn96}. The second measure was used in Ref. \cite{franchini08,its10}
to explore the entanglement of the anisotropic XY spin chain with a transverse magnetic field in the various phases. In order to compute the entanglement measures in our system we compute $\mbox{Tr} \rho_A^n$ up to ${\cal O} (\alpha)$ in section IV. 
In section V we compute the entanglement of formation ${\cal E}_{EoF} (\rho_0)$ and the second entanglement measure ${\cal E}_{\gamma} (\rho_0)$ within the first order of $\alpha$ when $\gamma$ is positive integer.
In this section it is shown that ${\cal E}_{\gamma} (\rho_0)$ increases with increasing $\alpha$ when $\gamma = 2, 3, \cdots$. However, it is also shown that the first-order term of $\alpha$ in ${\cal E}_{EoF} (\rho_0)$ is exactly zero. 
In section VI a brief conclusion is given. 

\section{Hamiltonian}

Let us consider the two coupled harmonic oscillator system, whose Hamiltonian is 
\begin{equation}
\label{hamiltonian-1}
\widehat{H}_2 = \frac{1}{2 m} \left( \widehat{P}_1^2 + \widehat{P}_2^2 \right) + \frac{1}{2} \left[ k_0 \left(\widehat{X}_1^2 + \widehat{X}_2^2 \right) + J \left( \widehat{X}_1 - \widehat{X}_2 \right)^2 \right],
\end{equation}
where $\left( \widehat{X}_i, \widehat{P}_i \right)$ obeys the GUP (\ref{GUP-d-2}). If we set 
\begin{equation}
\label{hup-1} 
\widehat{X}_j = \widehat{x}_j    \hspace{1.0cm} \widehat{P}_j = \widehat{p}_j (1 + \alpha \widehat{p}_j^2 )  \hspace{.5cm} (j = 1, 2),
\end{equation}
where $\left( \widehat{x}_j, \widehat{p}_j \right)$ obeys the HUP, $\widehat{H}_2$ becomes
\begin{equation}
\label{hamiltonian-2}
\widehat{H}_2 = \widehat{h}_1 + \widehat{h}_2 + \frac{J}{2} (\widehat{x}_1 - \widehat{x}_2)^2 + {\cal O} (\alpha^2)
\end{equation}
where 
\begin{equation}
\label{hamiltonian-3}
\widehat{h}_j = \frac{1}{2 m} \left( \widehat{p}_j^2 + 2 \alpha \widehat{p}_j^4 \right) + \frac{1}{2} k_0 \widehat{x}_j^2  \hspace{.5cm} (j= 1, 2).
\end{equation}

Now, we introduce the new coordinates
\begin{equation}
\label{normal-1}
\widehat{y}_1 = \frac{1}{\sqrt{2}} (\widehat{x}_1 + \widehat{x}_2)    \hspace{1.5cm}  \widehat{y}_2 = \frac{1}{\sqrt{2}} (-\widehat{x}_1 + \widehat{x}_2).
\end{equation}
Then, the Hamiltonian $\widehat{H}_2$ reduces to 
\begin{equation}
\label{hamitonian-4}
\widehat{H}_2 = \widehat{H}_0 + \Delta  \widehat{H}
\end{equation}
where  
\begin{equation}
\label{hamiltonian-5}
\widehat{H}_0 = \sum_{j=1}^2 \left[ \frac{1}{2m} \left( \widehat{\pi}_j^2 + \alpha \widehat{\pi}_j^4 \right) + \frac{1}{2} m \omega_j^2 \widehat{y}_j^2  \right] 
+ {\cal O} (\alpha^2)
\hspace{1.0cm} 
\Delta \widehat{H} = \frac{3 \alpha}{m} \widehat{\pi}_1^2 \widehat{\pi}_2^2.
\end{equation}
In eq. (\ref{hamiltonian-5}) $\widehat{\pi}_1$ and $\widehat{\pi}_2$ are the canonical momenta of $\widehat{y}_1$ and $\widehat{y}_2$, and the frequencies are 
\begin{equation}
\label{freq-1}
\omega_1 = \sqrt{\frac{k_0}{m}}     \hspace{1.0cm}    \omega_2 = \sqrt{\frac{k_0 + 2 J}{m}}.
\end{equation}
In next section we will derive the ground state for $\widehat{H}_2$ up to the order of $\alpha$ by treating $\Delta \widehat{H}$ as a small 
perturbation.

\section{Ground and its reduced states for $\widehat{H}_2$}

Before we solve the Schr\"{o}dinger equation for $\widehat{H}_2$, let us consider the one oscillator problem, whose Hamiltonian is 
$\widehat{H}_1 = \frac{\widehat{p}^2}{2 m} + \frac{\alpha}{m} \widehat{p}^4 + \frac{1}{2} m \omega^2 \widehat{x}^2 + {\cal O} (\alpha^2)$. In Ref. \cite{comment-1} the 
Schr\"{o}dinger equation for $\widehat{H}_1$ is solved up to ${\cal O} (\alpha)$. For example, the $n$th eigenstate and the corresponding eigenvalue are 
\begin{eqnarray}
\label{k-schrodinger-3}
&&\psi_n (x : \alpha, \omega)  = \phi_n (x : \omega)                                                                                                         \\    \nonumber
&& +                                                                                                       
 (\alpha m \hbar \omega)    \bigg[ \frac{(2 n + 3) \sqrt{(n+1) (n + 2)}}{4} \phi_{n + 2} (x : \omega) - \frac{(2 n - 1) \sqrt{n (n - 1)}}{4} \phi_{n - 2} (x : \omega)   \\  \nonumber
&&   +  \frac{\sqrt{n (n - 1) (n - 2) (n - 3)}}{16} \phi_{n-4} (x : \omega) - \frac{\sqrt{ (n + 1) (n + 2) (n + 3) (n + 4)}}{16} \phi_{n+4} (x : \omega)    \bigg] + {\cal O} (\alpha^2)    \\    \nonumber
&& {\cal E}_n (\alpha, \omega) = \left(n + \frac{1}{2} \right) \hbar \omega \left[ 1 + \frac{3 (2 n^2 + 2 n + 1)}{2 (2 n + 1)} (\alpha m \hbar \omega) \right] + {\cal O} (\alpha^2),
\end{eqnarray}
where $n = 0, 1, 2, \cdots$ and 
\begin{equation}
\label{k-schrodinger-4}
\phi_n (x: \omega) = \frac{1}{\sqrt{2^n n!}}  \left( \frac{m \omega}{\pi \hbar} \right)^{1/4} H_n \left( \sqrt{\frac{m \omega}{\hbar}} x \right) \exp \left[ - \frac{m \omega}{2 \hbar} x^2 \right].
\end{equation}
In Eq. (\ref{k-schrodinger-4}) $H_n (z)$ is a $n$th-order Hermite polynomial. We assume $\phi_m (z : \omega) = 0$ for $m < 0$. 

Now, let us consider the Schr\"{o}dinger equation for $\widehat{H}_0$:
\begin{equation}
\label{b-schro-1}
\widehat{H}_0 \phi_{n_1, n_2}^{(0)} (x_1, x_2) = E_{n_1, n_2}^{(0)} \phi_{n_1, n_2}^{(0)} (x_1, x_2).
\end{equation} 
Since $\widehat{H}_0$ is diagonalized, the eigenvalue and the corresponding eigenfunction are 
\begin{eqnarray}
\label{b-schro-2}
&&  E_{n_1, n_2}^{(0)} = {\cal E}_{n_1} \left( \frac{\alpha}{2}, \omega_1 \right) + {\cal E}_{n_2} \left( \frac{\alpha}{2}, \omega_2 \right)                  \\     \nonumber
&&\phi_{n_1, n_2}^{(0)} (x_1, x_2) = \psi_{n_1} \left(y_1 : \frac{\alpha}{2}, \omega_1 \right) \psi_{n_2} \left(y_2 : \frac{\alpha}{2}, \omega_2 \right).
\end{eqnarray} 

If we treat $\Delta \widehat{H}$ as a small perturbation, the ground state $\Phi_{0,0}$ and its eigenvalue $E_{0,0}$ for $\widehat{H}_2$ become
\begin{eqnarray}
\label{b-schro-3}
&&E_{0,0} = \frac{\hbar}{2} (\omega_1 + \omega_2) + \frac{3}{8} (\alpha m \hbar^2) (\omega_1 + \omega_2)^2 + {\cal O} (\alpha^2)       \\     \nonumber
&&\Phi_{0,0} (x_1, x_2) = \phi_0 (y_1 : \omega_1) \phi_0 (y_2 : \omega_2)                                                              \\     \nonumber
&& + (\alpha m \hbar) \Bigg[ \frac{3 \sqrt{2}}{8} (\omega_1 + \omega_2)  \left\{ \phi_0 (y_1 : \omega_1) \phi_2 (y_2 : \omega_2) + \phi_2 (y_1 : \omega_1) \phi_0 (y_2 : \omega_2) \right\}    \\    \nonumber
&&\hspace{2.0cm}  - \frac{\sqrt{6}}{16} \left\{ \omega_1 \phi_4 (y_1 : \omega_1) \phi_0 (y_2 : \omega_2) + \omega_2 \phi_0 (y_1 : \omega_1) \phi_4 (y_2 : \omega_2)   \right\}                 \\    \nonumber
&&\hspace{5.0cm}   - \frac{3}{4} \frac{\omega_1 \omega_2}{\omega_1 + \omega_2} \phi_2 (y_1 : \omega_1) \phi_2 (y_2 : \omega_2)    \Bigg]   + {\cal O} (\alpha^2).
\end{eqnarray}
Thus, the density matrix for the ground state is 
\begin{eqnarray}
\label{b-schro-4}
&&\rho_0 [x_1, x_2: x_1', x_2']                                                                                                     \\    \nonumber
&&= \Phi_{0,0} (x_1, x_2)  \Phi_{0,0}^* (x_1', x_2')                                                                                \\    \nonumber
&&= \phi_0 (y_1 : \omega_1)  \phi_0 (y_2, \omega_2)  \phi_0 (y_1' : \omega_1)  \phi_0 (y_2' : \omega_2)                             \\    \nonumber
&& + (\alpha m \hbar) \Bigg\{ \phi_0 (y_1 : \omega_1)  \phi_0 (y_2, \omega_2) \Bigg[ \frac{3 \sqrt{2}}{8} (\omega_1 + \omega_2)  \left\{ \phi_0 (y_1' : \omega_1) \phi_2 (y_2' : \omega_2) + \phi_2 (y_1' : \omega_1) \phi_0 (y_2' : \omega_2) \right\}    \\    \nonumber
&&\hspace{5.5cm}  - \frac{\sqrt{6}}{16} \left\{ \omega_1 \phi_4 (y_1' : \omega_1) \phi_0 (y_2' : \omega_2) + \omega_2 \phi_0 (y_1' : \omega_1) \phi_4 (y_2' : \omega_2)   \right\}                 \\    \nonumber
&&\hspace{7.0cm}   - \frac{3}{4} \frac{\omega_1 \omega_2}{\omega_1 + \omega_2} \phi_2 (y_1' : \omega_1) \phi_2 (y_2' : \omega_2)    \Bigg]                                                         \\     \nonumber
&& \hspace{8.0cm}              + (y_j \leftrightarrow y_j')  \Bigg\}  + {\cal O} (\alpha^2)
\end{eqnarray}
where
\begin{equation}
\label{normal-2}
y_1' = \frac{1}{\sqrt{2}} (x_1' + x_2')     \hspace{1.0cm}   y_2' = \frac{1}{\sqrt{2}} (-x_1' + x_2').
\end{equation}

In order to examine the entanglement of $\rho_0$ we should derive its reduced substate. After long and 
tedious calculation one can derive the reduced state $\rho_A = \mbox{Tr}_B \rho_0$ in a form
\begin{eqnarray}
\label{reduced-1}
&& \rho_A [ x_1, x_1']                                                  \\    \nonumber
&&\equiv \int dx_2 \rho_0 [x_1, x_2 : x_1', x_2]                        \\    \nonumber
&& = \sqrt{\frac{2 m \omega_1 \omega_2}{\pi \hbar (\omega_1 + \omega_2)}} e^{-a (x_1^2 + x_1'^2) + 2 b x_1 x_1'}  
\Bigg[ 1 + \frac{\alpha m}{256 \hbar (\omega_1 + \omega_2)^5} 
\Bigg\{ g_1 (x_1^4 + x_1'^4) + g_2 (x_1^3 x_1' + x_1 x_1'^3)                                    \\    \nonumber
&&\hspace{7.0cm}   + g_3 x_1^2 x_1'^2 + g_4 (x_1^2 + x_1'^2) + g_5 x_1 x_1' + g_6 \Bigg\}   \Bigg] + {\cal O} (\alpha^2)
\end{eqnarray}
where
\begin{equation}
\label{reduced-2}
a = \frac{m}{8 \hbar (\omega_1 + \omega_2)} (\omega_1^2 + \omega_2^2 + 6 \omega_1 \omega_2)    \hspace{1.0cm}
b = \frac{m (\omega_1 - \omega_2)^2}{8 \hbar (\omega_1 + \omega_2)}
\end{equation}
and
\begin{eqnarray}
\label{reduced-3}
&&g_1 = -m^2 (\omega_1^8 + 5 \omega_1^7 \omega_2 + 94 \omega_1^6 \omega_2^2 + 459 \omega_1^5 \omega_2^3 + 930 \omega_1^4 \omega_2^4 + 
459 \omega_1^3 \omega_2^5 + 94 \omega_1^2 \omega_2^6 + 5 \omega_1 \omega_2^7 + \omega_2^8)      \nonumber     \\
&&g_2 = 4 m^2 (\omega_1 - \omega_2)^2 (\omega_1^6 + 7 \omega_1^5 \omega_2 + 35 \omega_1^4 \omega_2^2 + 106 \omega_1^3 \omega_2^3 
+ 35 \omega_1^2 \omega_2^4 + 7 \omega_1 \omega_2^5 + \omega_2^6)                                \nonumber      \\
&&g_3 = -6 m^2 (\omega_1 - \omega_2)^4 (\omega_1^4 + 9 \omega_1^3 \omega_2 + 28 \omega_1^2 \omega_2^2 + 9 \omega_1 \omega_2^3 + 
\omega_2^4)                                                                                      \\      \nonumber
&& g_4 = 24 \hbar m (\omega_1 + \omega_2) (2 \omega_1^6 + 23 \omega_1^5 \omega_2 + 82 \omega_1^4 \omega_2^2 
+ 170 \omega_1^3 \omega_2^3 + 82 \omega_1^2 \omega_2^4 + 23 \omega_1 \omega_2^5 + 2 \omega_2^6)   \\     \nonumber
&&g_5 = -48 \hbar m (\omega_1 + \omega_2) (\omega_1 - \omega_2)^2 (2 \omega_1^4 + 11 \omega_1^3 \omega_2 + 30 \omega_1^2 \omega_2^2 
+ 11 \omega_1 \omega_2^3 + 2 \omega_2^4)                                                          \\     \nonumber
&&g_6 = -48 \hbar^2 (\omega_1 + \omega_2)^2 (4 \omega_1^4 + 17 \omega_1^3 \omega_2 + 38 \omega_1^2 \omega_2^2 + 
17 \omega_1 \omega_2^3 + 4 \omega_2^4).
\end{eqnarray}
It is useful to note 
\begin{equation}
\label{reduced-4}
a + b = \frac{m (\omega_1 + \omega_2)}{4 \hbar}    \hspace{1.0cm} a - b = \frac{m \omega_1 \omega_2}{\hbar (\omega_1 + \omega_2)}.
\end{equation}
Also, one can show
\begin{equation}
\label{reduced-5}
\frac{3}{16 (a - b)^2} (2 g_1 + 2 g_2 + g_3) + \frac{1}{4 (a - b)} (2 g_4 + g_5) + g_6 = 0.
\end{equation}
Using Eq. (\ref{reduced-5}) one can explicitly show $\mbox{Tr} \rho_A = 1$ within the leading order of $\alpha$, which 
guarantees that $\rho_A$ is a mixed quantum state. 

\begin{figure}[ht!]
\begin{center}
\includegraphics[height=6.0cm]{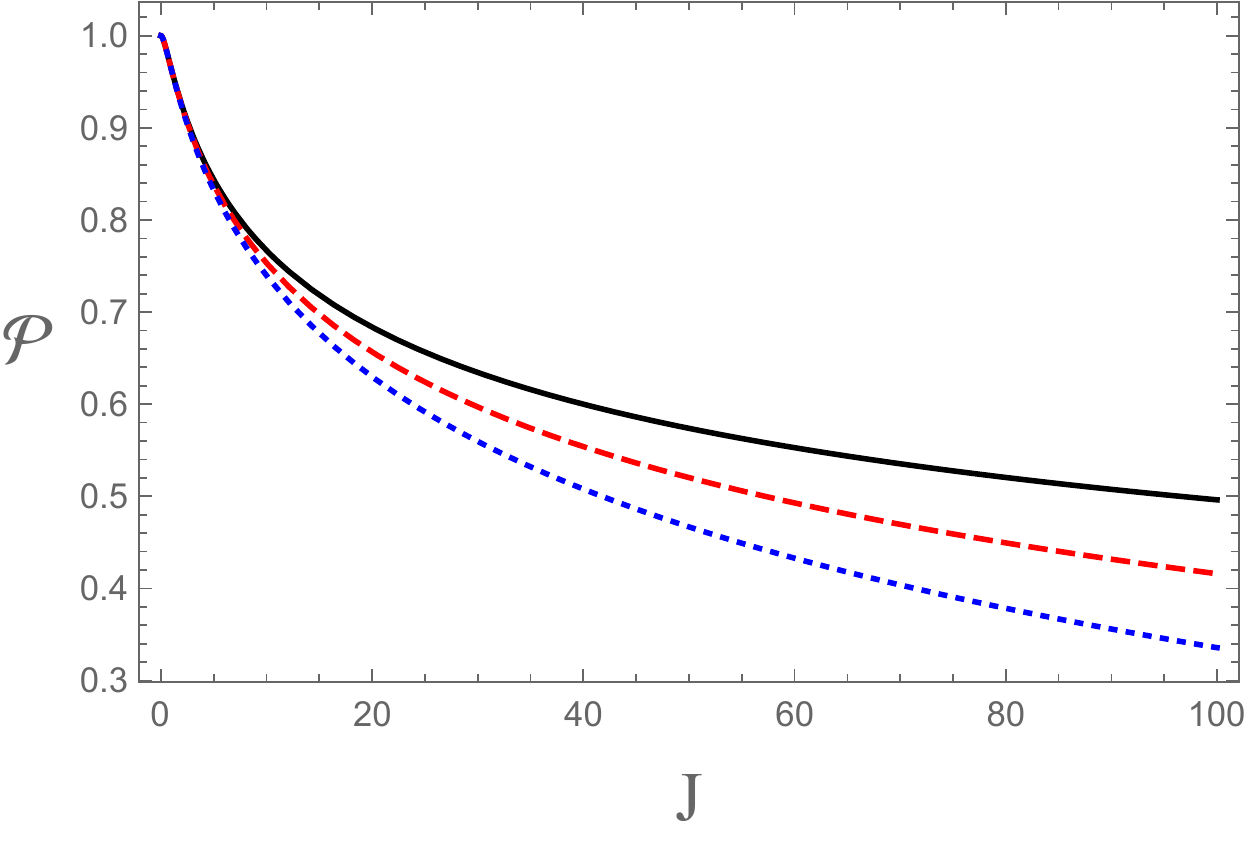} 

\caption[fig1]{(Color online) The $J$-dependence of the purity function when $k_0 = m = \hbar = 1$.  The black solid, red dashed, and blue dotted lines correspond to $\alpha = 0$, $0.2$, and $0.4$ respectively.
The figure shows that the reduced state $\rho_A$ becomes more mixed with increasing the GUP parameter $\alpha$.
 }
\end{center}
\end{figure}

In order to quantify how much $\rho_A$ is mixed we compute the purity function, whose expression is 
\begin{equation}
\label{purity-1}
{\cal P} (\rho_A) \equiv \mbox{Tr} \rho_A^2 
= \frac{2 \sqrt{\omega_1 \omega_2}}{\omega_1 + \omega_2} \left[ 1 - \frac{3 \alpha m \hbar}{32} \frac{(\omega_1 - \omega_2)^4}{(\omega_1 + \omega_2)^3} + {\cal O} (\alpha^2) \right].
\end{equation}
Fig. $1$ shows the $J$-dependence of the purity function when $k_0 = m = \hbar = 1$. The black solid, red dashed, and blue dotted lines correspond to $\alpha = 0$, $0.2$, and $0.4$ respectively. 
When $J=0$, $\rho_A$ is a pure state regardless of $\alpha$. With increasing $J$, $\rho_A$ becomes more and more mixed. The remarkable fact is that at fixed $J$ the GUP parameter $\alpha$ makes $\rho_A$ to be 
more mixed. This is due to the minus sign in the bracket of Eq. (\ref{purity-1}). 


\section{calculation of $\mbox{Tr} \rho_A^n$}

The most typical way for computing the R\'{e}nyi and von Neumann entropies of $\rho_A$ is to solve the eigenvalue equation 
\begin{equation}
\label{eigen-ent-1}
\int dx_1' \rho_A [ x_1, x_1'] f_n (x_1') = \lambda_n f_n (x_1).
\end{equation}
If $\rho_A$ is a Gaussian state, the eigenvalue equation (\ref{eigen-ent-1}) can be solved straightforwardly\cite{serafini}.
Then, the R\'{e}nyi entropy of order $\gamma$ and von Neumann entropy can be computed by making use of the eigenvalue $\lambda_n$ as following:
\begin{equation}
\label{eigen-ent-2}
{\cal E}_{\gamma} (\rho_0) = \frac{1}{1 - \gamma} \ln \sum_n \left( \lambda_n \right)^{\gamma}       \hspace{1.0cm}  {\cal E}_{EoF} (\rho_0) = - \sum_n \lambda_n \ln \lambda_n,
\end{equation}
where $\gamma$ is arbitrary nonnegative real.
The problem is that $\rho_A$ is not Gaussian state if $\alpha \neq 0$ as Eq. (\ref{reduced-1}) shows. Thus, it seems to be extremely difficult to solve 
Eq. (\ref{eigen-ent-1}) directly. 

Although we cannot solve the eigenvalue equation (\ref{eigen-ent-1}) explicitly, we can compute ${\cal E}_{\gamma = n} (\rho_0)$ and ${\cal E}_{EoF} (\rho_0)$ at least up to the ${\cal O} (\alpha)$ by computing $\mbox{Tr} \rho_A^n$ \cite{gup-sho}.
In this case ${\cal E}_{\gamma = n} (\rho_0)$ can be computed by 
\begin{equation}
\label{eigen-ent-3}
{\cal E}_{\gamma = n} (\rho_0) = \frac{1}{1 - n} \ln \mbox{Tr} \rho_A^n. 
\end{equation}
Then, ${\cal E}_{EoF} (\rho_0)$ also can be computed from Eq. (\ref{eigen-ent-3}) by taking $n \rightarrow 1$ limit. 
In this reason we will compute $\mbox{Tr} \rho_A^n$ in this section within ${\cal O} (\alpha)$. 

From Eq. (\ref{reduced-1}) one can show 
\begin{eqnarray}
\label{tracen-1}
&&\mbox{Tr} \rho_A^n                                                                                                                                                            
\equiv \int dx_1 \cdots dx_n \rho_A[x_1, x_2] \rho_A[x_2, x_3] \cdots \rho_A[x_{n-1}, x_n] \rho_A[x_n, x_1]            \\    \nonumber
&&\hspace{1.0cm} = \left(\frac{2 m \omega_1 \omega_2}{\pi \hbar (\omega_1 + \omega_2)} \right)^{n/2}   \int  dx_1 \cdots dx_n \exp \left[ -{\bm X} G_n {\bm X}^{\dagger} \right]    \\    \nonumber
&& \times \Bigg\{ 1 + \frac{\alpha m}{256 \hbar (\omega_1 + \omega_2)^5} \Bigg[ 2 g_1 \left( x_1^4 + \dots + x_n^4 \right)                                              \\    \nonumber
&&\hspace{4.5cm} + g_2 \left[ x_1 x_2 (x_1^2 + x_2^2) + \cdots + x_{n-1} x_n (x_{n-1}^2 + x_n^2) + x_n x_1 (x_n^2 + x_1^2) \right]                                                     \\    \nonumber
&&\hspace{5cm}+ g_3 (x_1^2 x_2^2 + \cdots + x_{n-1}^2 x_n^2 + x_n^2 x_1^2 )  + 2 g_4 (x_1^2 + \cdots + x_n^2)                                              \\    \nonumber
&&\hspace{5cm}+ g_5 (x_1 x_2 + \cdots + x_{n-1} x_n + x_n x_1) + n g_6    \Bigg] + {\cal O} (\alpha^2)     \Bigg\},
\end{eqnarray}
where $X$ is a $n$-dimensional row vector defined by ${\bm X} = (x_1, x_2, \cdots, x_n)$ and $G_n$ is a $n \times n$ matrix given by
\begin{eqnarray}
\label{tracen-2}
G_n = \left(                  \begin{array}{cccccc}
                  2 a & \hspace{0.2cm}    -b    & \hspace{0.2cm}         & \hspace{0.2cm}         & \hspace{0.2cm}         & \hspace{0.2cm}  -b            \\
                  -b  & \hspace{0.2cm} 2 a      & \hspace{0.2cm} \bullet & \hspace{0.2cm}         & \hspace{0.2cm}         & \hspace{0.2cm}                \\
                      & \hspace{0.2cm} \bullet  & \hspace{0.2cm} \bullet & \hspace{0.2cm} \bullet & \hspace{0.2cm}         & \hspace{0.2cm}                \\
                      & \hspace{0.2cm}          & \hspace{0.2cm} \bullet & \hspace{0.2cm} \bullet & \hspace{0.2cm} \bullet & \hspace{0.2cm}                \\
                      & \hspace{0.2cm}          & \hspace{0.2cm}         & \hspace{0.2cm} \bullet & \hspace{0.2cm} \bullet & \hspace{0.2cm}   -b           \\
                   -b   & \hspace{0.2cm}          & \hspace{0.2cm}         & \hspace{0.2cm}         & \hspace{0.2cm}     -b  & \hspace{0.2cm}  2a 
                              \end{array}                        \right).
\end{eqnarray} 
In Eq. (\ref{tracen-2}) the matrix components in the empty space are all zero. As shown in Ref. \cite{gup-sho}, the determinant of $G_n$ is 
\begin{equation}
\label{tracen-3}
\det G_n = 2^{-n} \left[ \left( \sqrt{a + b} + \sqrt{a - b} \right)^n -  \left( \sqrt{a + b} - \sqrt{a - b} \right)^n \right]^2 = \left( \frac{m}{8 \hbar (\omega_1 + \omega_2)} \right)^n {\cal Z}_{-, 2n}^2
\end{equation}
where 
\begin{equation}
\label{tracen-4} 
{\cal Z}_{\pm, \ell} = \left( \sqrt{\omega_2} + \sqrt{\omega_1} \right)^{\ell} \pm \left( \sqrt{\omega_2} - \sqrt{\omega_1} \right)^{\ell}.
\end{equation}
Then, it is possible to show 
\begin{eqnarray}
\label{tracen-5}
&&\int dx_1 \cdots dx_n  \exp \left[- {\bm X} G_n {\bm X}^{\dagger} \right] = \frac{\pi^{n/2}}{\sqrt{\det G_n}} \equiv h_n                                 \\     \nonumber
&&\int dx_1 \cdots dx_n (x_1^2 + \cdots + x_n^2) \exp \left[- {\bm X} G_n {\bm X}^{\dagger} \right]                                                              \\     \nonumber
&&= h_n \frac{n}{4 \sqrt{a^2 - b^2}} \frac{(\sqrt{a + b} + \sqrt{a - b})^n + (\sqrt{a + b} - \sqrt{a - b})^n}
                                                                                          {(\sqrt{a + b} + \sqrt{a - b})^n - (\sqrt{a + b} - \sqrt{a - b})^n}    \\     \nonumber
&&= h_n \frac{n \hbar}{2 m \sqrt{\omega_1 \omega_2}} \frac{{\cal Z}_{+,2n}}{{\cal Z}_{-,2n}}     \\    \nonumber
&&\int dx_1 \cdots dx_n (x_1 x_2 + \cdots + x_{n-1} x_n + x_n x_1)  \exp \left[- {\bm X} G_n {\bm X}^{\dagger} \right]                                           \\     \nonumber
&&= h_n \frac{n b}{2 \sqrt{a^2 - b^2}} \frac{(\sqrt{a + b} + \sqrt{a - b})^{n-2} + (\sqrt{a + b} - \sqrt{a - b})^{n-2}}
                                                                                           {(\sqrt{a + b} + \sqrt{a - b})^n - (\sqrt{a + b} - \sqrt{a - b})^n}    \\     \nonumber
&&= h_n \frac{n \hbar (\omega_1 - \omega_2)^2}{2 m \sqrt{\omega_1 \omega_2}}  \frac{{\cal Z}_{+, 2n-4}}{{\cal Z}_{-,2n}}.
\end{eqnarray}
Also, one can show\cite{gup-sho}
\begin{eqnarray}
\label{tracen-6}
&&\int dx_1 \cdots dx_n (x_1^4 + \cdots + x_n^4)  \exp \left[- {\bm X} G_n {\bm X}^{\dagger} \right]                                                                    \\   \nonumber
&&= h_n \frac{3 n}{4} \frac{(\mbox{det} H_{n-1})^2}{(\det G_n)^2} 
= h_n \frac{3 n \hbar^2}{4 m^2 \omega_1 \omega_2} \left( \frac{{\cal Z}_{+,2n}}{{\cal Z}_{-,2n}} \right)^2                                                              \\   \nonumber
&&\int dx_1 \cdots dx_n (x_1^2 x_2^2 + \cdots + x_{n-1}^2 x_n^2 + x_n^2 x_1^2) \exp \left[- {\bm X} G_n {\bm X}^{\dagger} \right]                                       \\    \nonumber
&&= h_n \frac{n}{4 (\det G_n)^2} \Bigg[ 12 a^2 (\det H_{n-2})^2 - 12 a b^2 (\det H_{n-2}) (\det H_{n-3})                                                                 \\    \nonumber
&& \hspace{5.0cm}  + 3 b^4 (\det H_{n-3})^2 - 2 (\det H_{n-2}) (\det G_n) \bigg]                                                                                         \\   \nonumber  
&& = h_n \frac{n \hbar^2}{4 m^2 \omega_1 \omega_2 {\cal Z}_{-,2n}^2}   \bigg[ 3 {\cal Z}_{+,2n}^2 - 16 (\omega_1 + \omega_2) \sqrt{\omega_1 \omega_2} {\cal Z}_{-, 4n-4} \bigg]     \\   \nonumber                                                                                                                                           
&&\int dx_1 \cdots dx_n \left[ x_1 x_2 (x_1^2 + x_2^2) + \cdots + x_{n-1} x_n (x_{n-1}^2 + x_n^2) + x_n x_1 (x_n^2 + x_1^2) \right]                                       \\    \nonumber
&& \hspace{8.0cm}  \times \exp \left[- {\bm X} G_n {\bm X}^{\dagger} \right]                                                                                               \\    \nonumber
&&= h_n \frac{3 n}{2} \frac{ \left[ b^{n-1} + b (\det H_{n-2}) \right] \left[ 2 a (\det H_{n - 2}) - b^2 (\det H_{n - 3}) \right]}{(\det G_n)^2}                            \\   \nonumber
&&= h_n \frac{3 n \hbar^2 (\omega_1 - \omega_2)^2}{2 m^2 \omega_1 \omega_2}  \frac{{\cal Z}_{+,2n}}{{\cal Z}_{-,2n}^3} \bigg[ 8 (\omega_1 + \omega_2) \sqrt{\omega_1 \omega_2} (\omega_1 - \omega_2)^{2(n-2)} + {\cal Z}_{-,4n-4} \bigg],
\end{eqnarray}
where $H_n$ is a $n \times n$ tridiagonal matrix given by 
\begin{eqnarray}
\label{tracen-7}
H_n = \left(                  \begin{array}{cccccc}
                  2 a & \hspace{0.2cm}    -b    & \hspace{0.2cm}         & \hspace{0.2cm}         & \hspace{0.2cm}         & \hspace{0.2cm}                \\
                  -b  & \hspace{0.2cm} 2 a      & \hspace{0.2cm} \bullet & \hspace{0.2cm}         & \hspace{0.2cm}         & \hspace{0.2cm}                \\
                      & \hspace{0.2cm} \bullet  & \hspace{0.2cm} \bullet & \hspace{0.2cm} \bullet & \hspace{0.2cm}         & \hspace{0.2cm}                \\
                      & \hspace{0.2cm}          & \hspace{0.2cm} \bullet & \hspace{0.2cm} \bullet & \hspace{0.2cm} \bullet & \hspace{0.2cm}                \\
                      & \hspace{0.2cm}          & \hspace{0.2cm}         & \hspace{0.2cm} \bullet & \hspace{0.2cm} \bullet & \hspace{0.2cm}   -b           \\
                      & \hspace{0.2cm}          & \hspace{0.2cm}         & \hspace{0.2cm}         & \hspace{0.2cm}     -b  & \hspace{0.2cm}  2a 
                              \end{array}                        \right).
\end{eqnarray}
It is straightforward to show 
\begin{eqnarray}
\label{tracen-8}
\det H_n &=& \frac{1}{\sqrt{a^2 - b^2}} \left[ a \sqrt{\det G_{2n}} - \frac{b^2}{2} \sqrt{\det G_{2n-2}} \right]                                     \\    \nonumber
&=& \frac{1}{8 (\omega_1 + \omega_2) \sqrt{\omega_1 \omega_2}} \left( \frac{m}{8 \hbar (\omega_1 + \omega_2)} \right)^n {\cal Z}_{-,4n+4}.
\end{eqnarray}
Eq. (\ref{tracen-8}) is valid for any nonnegative integer $n$.

Inserting Eqs. (\ref{tracen-5}) and (\ref{tracen-6}) into Eq (\ref{tracen-1}),  one can show that $\mbox{Tr} \rho_A^n$ can be written as a form 
\begin{equation}
\label{tracen-9}
\mbox{Tr} \rho_A^n = \frac{(1 - \xi)^n}{1 - \xi^n} \left[ 1 - \frac{3 n (\alpha m \hbar)}{4096} \frac{(\omega_1 - \omega_2)^4}{\omega_1 \omega_2 (\omega_1 + \omega_2)^5} \frac{{\cal Z}_{-,4}}{{\cal Z}_{-,2n}^2} {\cal J}_n (\omega_1, \omega_2) 
+ {\cal O} (\alpha^2) \right]
\end{equation}
where 
\begin{eqnarray}
\label{tracen-10}
&&{\cal J}_n (\omega_1, \omega_2) = {\cal Z}_{-,2n} \left[ {\cal Z}_{+,2n+4} + 2 (\omega_1 - \omega_2)^2 {\cal Z}_{+,2n} - 2 (\omega_1 - \omega_2)^6 {\cal Z}_{+,2n-8}  \right]        \\    \nonumber
&&   \hspace{3.0cm}  -3 (\omega_1 - \omega_2)^{2 n} {\cal Z}_{-,4} - (\omega_1 - \omega_2)^8 {\cal Z}_{-,4n-12}
\end{eqnarray}
and $\xi = \left[ (\sqrt{\omega_2} - \sqrt{\omega_1}) / (\sqrt{\omega_2} + \sqrt{\omega_1}) \right]^2$.

It is easy to show that when $n=2$, Eq. (\ref{tracen-9}) reproduces the purity function in Eq. (\ref{purity-1}). When $n=3$, Eq. (\ref{tracen-9}) yields
\begin{equation}
\label{tracen-11}
\mbox{Tr} \rho_A^3 = \frac{16 \omega_1 \omega_2}{(3 \omega_1 + \omega_2) (\omega_1 + 3 \omega_2)} 
\left[ 1 - \frac{9 (\alpha m \hbar)}{4} \frac{(\omega_1 + \omega_2) (\omega_1 - \omega_2)^4}{(3 \omega_1 + \omega_2)^2 (\omega_1 + 3 \omega_2)^2} + {\cal O} (\alpha^2)  \right].
\end{equation}
It is not difficult to show that as expected, Eq. (\ref{tracen-11}) exactly coincides with $\int dx_1 dx_2 dx_3 \rho_A[x_1, x_2] \rho_A [x_2, x_3] \rho_A [x_3, x_1]$. In next section we will discuss on the entanglement of $\rho_0$ by 
making use of Eq. (\ref{tracen-9}).

\section{Entanglement for $\rho_0$}

\begin{figure}[ht!]
\begin{center}
\includegraphics[height=6.0cm]{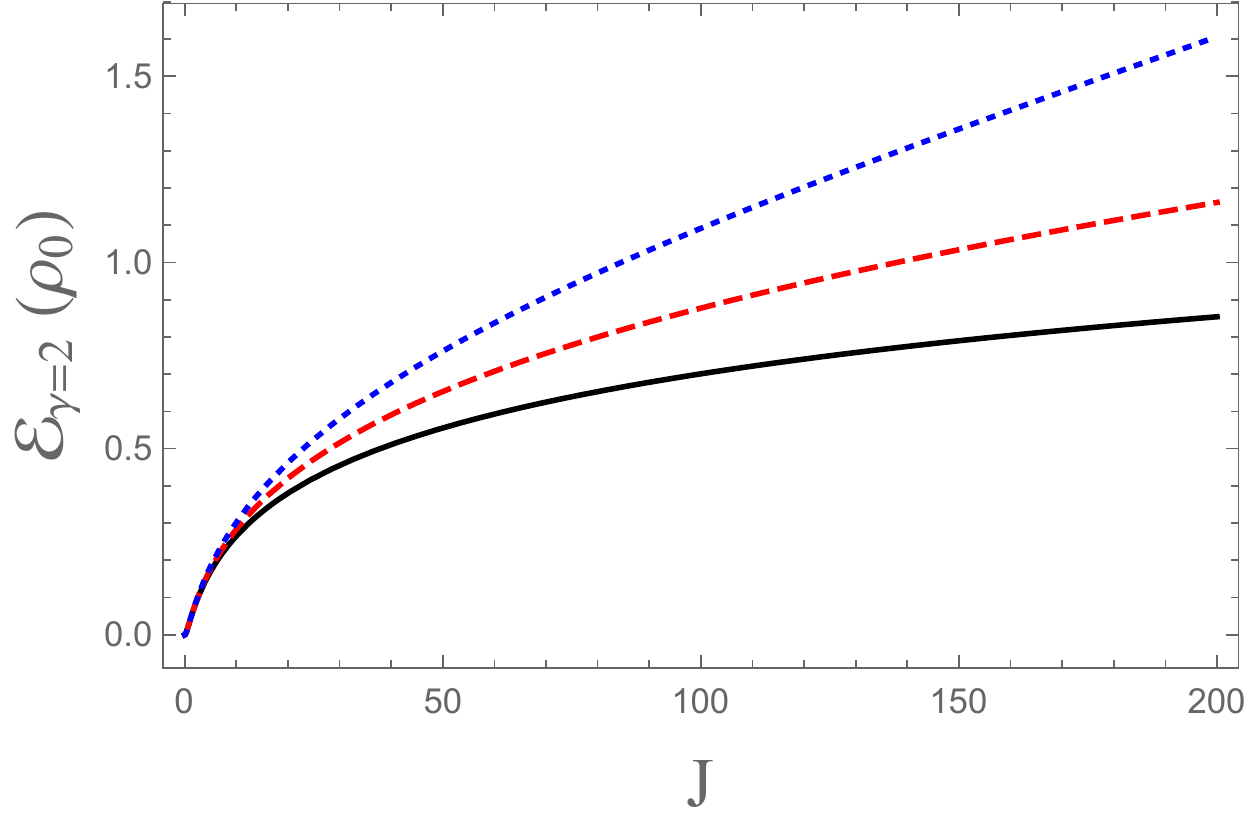} 

\caption[fig2]{(Color online)  The $J$-dependence of the ${\cal E}_{\gamma = 2} (\rho_0)$ when $k_0 = m = \hbar = 1$.  
The black solid, red dashed, and blue dotted lines correspond to $\alpha = 0$, $0.2$, and $0.4$ respectively.
This figure shows that it increases with increasing the GUP parameter $\alpha$.
 }
\end{center}
\end{figure}

The second entanglement measure ${\cal E}_{\gamma = n} (\rho_0)$ can be derived by inserting Eq. (\ref{tracen-9}) into Eq. (\ref{eigen-ent-3}), which is 
\begin{equation}
\label{renyi-1}
{\cal E}_{\gamma = n} = \frac{1}{1 - n} \left[ \ln \frac{(1 - \xi)^n}{1 - \xi^n} + \ln \left\{  1 - \frac{3 n (\alpha m \hbar)}{4096} \frac{(\omega_1 - \omega_2)^4}{\omega_1 \omega_2 (\omega_1 + \omega_2)^5} \frac{{\cal Z}_{-,4}}{{\cal Z}_{-,2n}^2} {\cal J}_n (\omega_1, \omega_2) + {\cal O} (\alpha^2) \right\} \right].
\end{equation}

Now, let us compute the EoF of $\rho_0$. This is achieved by taking  $n \rightarrow 1$ limit to Eq. (\ref{renyi-1}). One can show that ${\cal J}_n (\omega_1, \omega_2)$ in Eq. (\ref{tracen-10}) satisfies 
\begin{equation}
\label{eof-1}
{\cal J}_1 (\omega_1, \omega_2) = \frac{d}{d n} {\cal J}_n (\omega_1, \omega_2) \Bigg|_{n = 1} = 0.
\end{equation}
Eq. (\ref{eof-1}) implies that the EoF of $\rho_0$ does not involve the first order of $\alpha$. Thus, it is expressed as 
\begin{equation}
\label{eof-2}
{\cal E}_{EoF} (\rho_0) = - \ln (1 - \xi) - \frac{\xi}{1 - \xi} \ln \xi + {\cal O} (\alpha^2).
\end{equation}

In Fig. 2 we plot the $J$-dependence of ${\cal E}_{\gamma = 2} (\rho_0)$ when $\alpha$ is $0$ (black solid line), $0.2$ (red dashed line), and $0.4$ (blue dotted line). We set $k_0 = m = \hbar = 1$ for simplicity. 
This figure shows that
it increases with increasing the GUP parameter $\alpha$. This can be seen from the fact that the second term in the bracket of 
Eq. (\ref{renyi-1}) increases in the negative region with respect to $\alpha$.

\section{Conclusions}

\begin{figure}[ht!]
\begin{center}
\includegraphics[height=5.3cm]{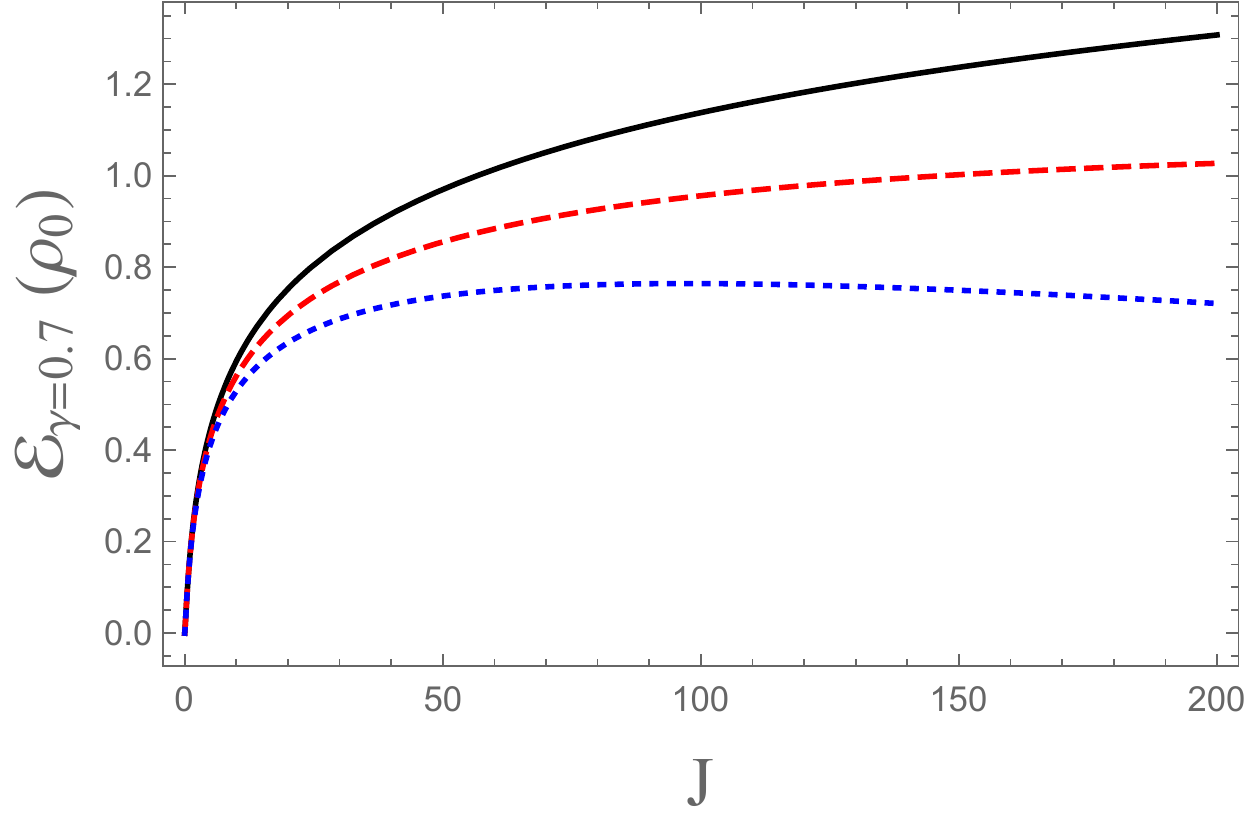} 
\includegraphics[height=5.3cm]{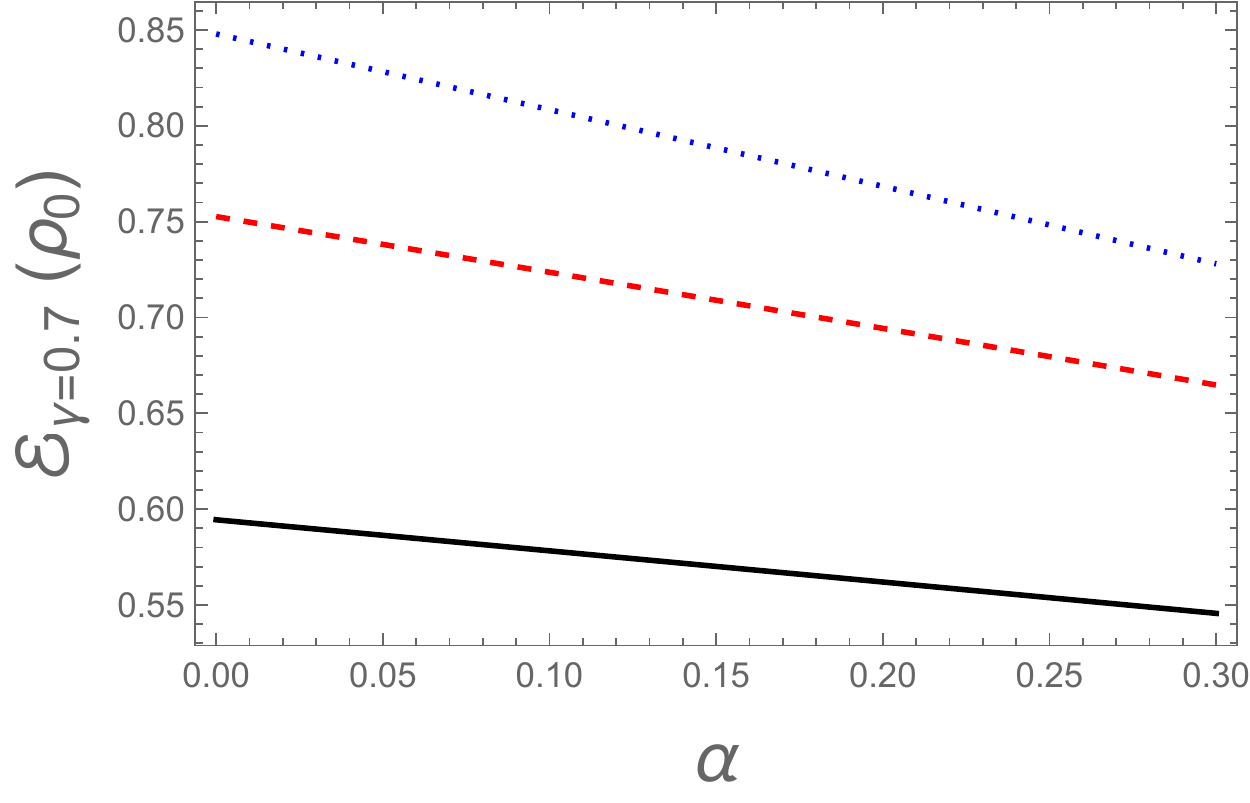} 

\caption[fig3]{(Color online)  (a) The $J$-dependence of the ${\cal E}_{\gamma = 0.7} (\rho_0)$ when $k_0 = m = \hbar = 1$.  
The black solid, red dashed, and blue dotted lines correspond to $\alpha = 0$, $0.2$, and $0.4$ respectively. (b) The $\alpha$-dependence of the ${\cal E}_{\gamma = 0.7} (\rho_0)$ when $k_0 = m = \hbar = 1$. The black solid, red dashed, and blue dotted lines correspond to $J = 10$, $20$, and $30$ respectively.
Both figures shows that it decreases with increasing the GUP parameter $\alpha$.}
\end{center}
\end{figure}

In this paper we examine how the quantum entanglement is modified in the GUP-corrected quantum mechanics. 
In order to explore this issue we consider the coupled harmonic oscillator system. Constructing the vacuum state $\rho_0$ and its substate $\rho_A$, we compute the entanglement
by choosing the EoF ${\cal E}_{EoF} (\rho_0) = S_{von} (\rho_A)$ and the R\'{e}nyi entropy  ${\cal E}_{\gamma} (\rho_0) = S_{\gamma} (\rho_A)$ of the substate as entanglement measures.
It is shown that the second entanglement measure increases with increasing $\alpha$ when $\gamma = 2, 3, \cdots$. Remarkable fact is that the EoF is invariant within the first-order of $\alpha$
in quantum mechanics with HUP and GUP.
Since ${\cal E}_{EoF} (\rho_0) = \lim_{\gamma \rightarrow 1} {\cal E}_{\gamma} (\rho_0)$, we conjecture that ${\cal E}_{\gamma} (\rho_0)$ decreases with increasing $\alpha$ when $\gamma < 1$.  

In order to compute ${\cal E}_{\gamma} (\rho_0)$ for nonnegative real $\gamma$ we should derive the eigenvalue $\lambda_n$ in Eq. (\ref{eigen-ent-1}).
However, it seems to be highly difficult (or might be impossible) to derive the eigenvalue due to non-Gaussian nature of $\rho_A$. Thus, we cannot confirm our conjecture directly. 
If ${\cal E}_{\gamma} (\rho_0)$ is equal to the right-hand side of Eq. (\ref{renyi-1}) with changing only $n \rightarrow \gamma$ for all nonnegative real $\gamma$, it is possible to show that our conjecture is right.
For example, we plot the $J$- dependence and $\alpha$-dependence of ${\cal E}_{\gamma = 0.7} (\rho_0)$ in Fig. 3(a) and Fig. 3(b) respectively.  
We choose various $\alpha$ in Fig. 3(a) and various J in Fig. 3(b). 
These two figures show that 
${\cal E}_{\gamma = 0.7} (\rho_0)$ decreases with increasing $\alpha$, which is consistent with our conjecture. 

The well-known example of the Planck scale is an early universe ($t \leq 10^{-43} (s)$ after big bang). However, we do not understand the role of quantum information at this early stage, in particular in the 
context of cosmology. We hope to explore this issue in the future. 

\vspace{1.0cm}

{\bf Acknowledgments}:
This work was supported by the National Research Foundation of Korea(NRF) grant funded by the Korea government(MSIT) (No. 2021R1A2C1094580).

\end{document}